\documentclass[amsmath,amssymb,aps,prl,tightenlines,twocolumn]{revtex4-1}

\usepackage[english]{babel}
\usepackage{afterpage}
\usepackage{graphicx}
\usepackage{dcolumn}
\usepackage{bm}
\usepackage{color}
\usepackage{slashed}
\usepackage{latexsym}

\begin{document}

\title{Laser-induced electron Fresnel diffraction in the tunneling and over-barrier ionization}

\author{Lei Geng$^1$}
\author{Hao Liang$^1$}
\author{Liang-You Peng$^{1,2,3,4}$}
\email{liangyou.peng@pku.edu.cn}

\affiliation{
$^1$State Key Laboratory for Mesoscopic Physics and Frontiers Science Center for Nano-optoelectronics, School of Physics, Peking University, 100871
Beijing, China\\
$^2$Collaborative Innovation Center of Quantum Matter, Beijing 100871, China\\
$^3$Collaborative Innovation Center of Extreme Optics, Shanxi University, 030006 Taiyuan, China \\
$^4$Beijing Academy of Quantum Information Sciences, Beijing 100193, China
}

\begin{abstract}
  The photoelectron momentum distribution in the strong-field ionization has a variety of structures that reveal the complicated dynamics of this process.  Recently, we identified a low-energy interference structure in the case of a super-intense extreme ultraviolet~(XUV) laser pulse and attributed it to the laser-induced electron Fresnel diffraction. This structure is determined by the laser-induced electron displacement~[Geng L {\it et al.} 2021 {\it Phys. Rev. A} {\bf 104}(2) L021102]. In the present work,  we find  that the Fresnel diffraction picture is also present in the tunneling and over-barrier regime of ionization by  short  pulses.  However, the electron displacement is now induced by the electric field component of the laser pulse, instead of  by the  magnetic field component in the case of the super-intense XUV pulse. After corresponding modifications to our quantum and semiclassical models, we find the same physical mechanism of the Fresnel diffraction governs the low-energy interference structures along the laser polarization.  The results predicted by the two models agree well with the  accurate results from the numerical solution to he time-dependent Schr\"odinger equation.
\end{abstract}
\maketitle

The laser is an effective tool   to investigate physics at the atomic level. Since the advent of the chirped-pulse amplification technique (CPA)~\cite{strickland1985}, many nonlinear phenomena such as high harmonics generation (HHG)~\cite{lewenstein1994,guo2021} and above threshold ionization (ATI)~\cite{corkum1989,zhang2017} have been observed and explained.   In recent years, advances in experimental techniques have enabled researchers to learn more about atomic ionization dynamics. Using the cold-target recoil-ion-momentum spectroscopy (COLTRIMS)~\cite{Ullrich1997} or the velocity map imaging (VMI) spectrometers~\cite{Eppink97}, one can obtain the photoelectron momentum distribution (PMD).  The PMD  contains a wealth of information regarding the process of ionization. Some structures in the PMD such as spider and fork structures have been explained as the interference of different trajectories in a semiclassical picture~\cite{Figueira_de_Morisson_Faria_2020}. At the same time, a more rigorous theoretical analysis can be carried out by numerically solving the time-dependent Schr\"odinger equation (TDSE)~\cite{Huismans2011,Jiang2020}. The combination of experiments and theories promotes the development of atomic physics in strong laser field.

In our previous work~\cite{Geng2021}, we focus on the ionization in the case of a super-intense XUV laser pulse, whose intensity reaches $10^{20}~{\rm W/{cm}^2}$. A novel low-energy interference structure is identified in the PMD along the laser propagation direction by solving the nondipole TDSE. When using the dipole approximation,  the novel structure would  disappear. According to our findings, the magnetic force causes the electron to move along the laser propagation direction and induce a displacement at the end of the laser pulse. Due to the diffusion of the electron wave packet, the electron will interact with the nucleus again, resulting in the interference structure. The last process is compared with the Fresnel diffraction in the wave optics, which occurs when the distance between the source and the obstruction is finite. Based on this picture, analytical quantum and semiclassical models have been developed.  Our models clearly show that  the details of the particular  interferences   are determined by the electron displacement induced by the laser.

In this Letter, TDSE is solved numerically for the hydrogen atom ionized by half-cycle pulses with substantially lower peak intensity ($10^{14}~{\rm W/{cm}^2} \sim 10^{16}~{\rm W/{cm}^2}$). The nondipole effect is negligible for such intensities, hence the dipole approximation is employed in these cases. The finite element discrete variable representation~\cite{McCurdy2000} and the Arnoldi propagator~\cite{Park1986} are used in our numerical methods. We find comparable petal-like patterns in the PMDs for half-cycle situations, in which laser settings are quite different from those of super-intense XUV cases. This is due to the fact that the underlying physical mechanisms  are similar. The Fresnel diffraction picture will be shown to be valid not only in the extreme condition like those in  Ref.~\cite{Geng2021}, but also in some tunneling and over-barrier (OB) cases. The Keldysh parameters $\gamma=\omega\sqrt{2I_p}/F$ are less than one in both tunneling and OB regimes. But the electric field is stronger in the OB regime. As a consequence, the electron will be ionized much faster in this situation. A common distinguishing criterion is given in Ref.~\cite{Morishita2013}. For the ground state of the hydrogen atom, the corresponding threshold intensity is about $5.5\times10^{14}~{\rm W/{cm}^2}$. We will concentrate on two half-cycle cases in the following sections, one in the OB regime and the other in the tunneling regime. In the case of OB ionization, the electron is ionized at the beginning of laser pulses, which is similar to the super-intense XUV scenario. For a better comparison, a specific form of the vector potential is adopted. The situation is a little different in the case of tunneling ionization. The electron is ionized at the peak of the electric field. And the site of the tunneling exit and the ionization rate must be taken into account in the model. Atomic units are employed throughout this Letter unless otherwise stated.

\vskip 4mm
\centerline{\includegraphics[width=\linewidth]{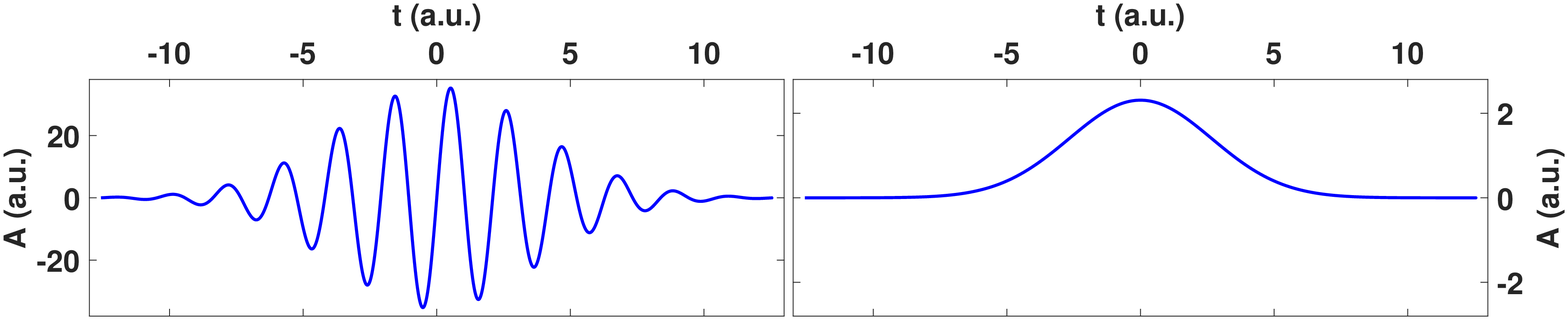}}\vspace{-0.8mm}
\centerline{\includegraphics[width=\linewidth]{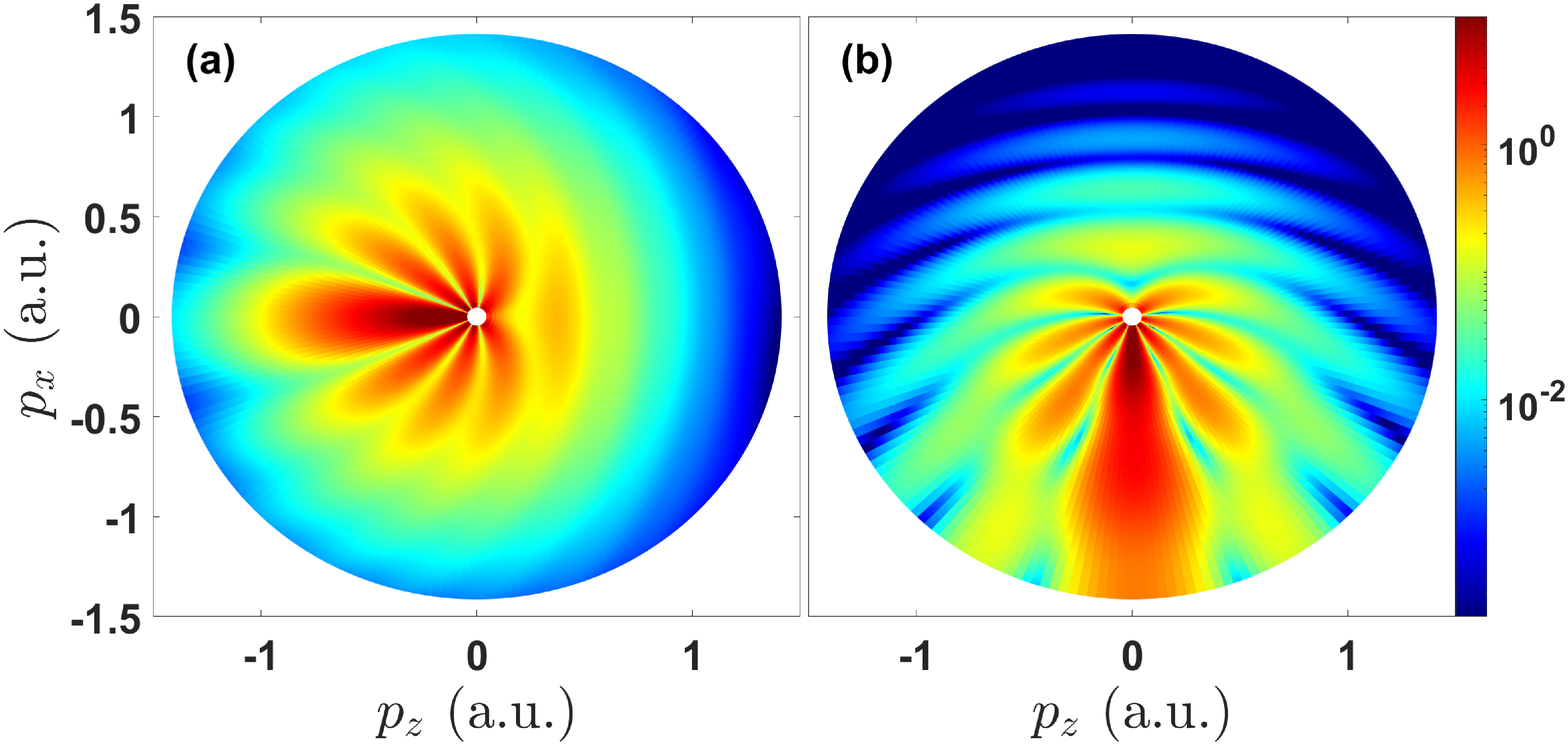}}
\vskip 2mm
\centerline{\footnotesize \begin{tabular}{p{7.5cm}}\bf Fig.\,1. \rm
    The PMDs in the $xz$-plane for XUV case~(a) and OB case~(b). The vector potentials of these two cases are plotted above the corresponding PMDs. Note that the scales for two laser pulses are different.
\end{tabular}}
\vskip 0.5\baselineskip

Let's pay attention to the OB case first. It will be compared with the XUV case. The vector potential and PMD of the case in Ref.~\cite{Geng2021} is presented on the left side of Fig.\,1. The XUV vector potential can be written as $\bm{A}_{\rm XUV}(t)=f(t) \sin(\omega t)~{\hat{\bm e}_x}$. For comparison, a particular half-cycle pulse that is associated with the XUV pulse is chosen in the example of OB ionization. Its vector potential is set to be $\bm{A}_{\rm OB}(t)=f(t)^2/2c~{\hat{\bm e}_x}$, where $c \approx 137~{\rm a.u.}$ is the speed of light. This means that its full width at half maximum (FWHM) is about $6.28~{\rm a.u.}$. We will explain why this form is adopted in the following. Note that the electric field is ${\bm F}(t)=-\mathop{}\!\mathrm{d}{\bm A}(t)/\mathop{}\!\mathrm{d} t$. One can get the peak laser intensity is about $9.7\times10^{15}~{\rm W/{cm}^2}$ for the parameters used in Ref.~\cite{Geng2021}, which is much lower than that of XUV case. The frequency of the half-cycle pulse is  not properly defined due to the wide frequency spectrum. In this case, an effective carrier frequency can be estimated by $F_{\rm max}/A_{\rm max}\approx0.228~{\rm a.u.}$. As a result, the effective Keldysh parameter is 0.433. The vector potential and PMD of the OB case are shown on the right side of Fig.\,1. It is intriguing that these two distinct cases, one in the OB regime and the other in the stabilization regime of laser-matter interactions~\cite{eberly1993}, have similar low-energy interference structures except for their orientations.

The explanation for the formation of the low-energy structures in the XUV situation has been examined. It demonstrates that the electron displacement induced by the magnetic field components of the laser, which is $\int_{-\infty}^{\infty}A_{\rm XUV}(t)^2/2c \mathop{}\!\mathrm{d} t~ {\hat{\bm e}_z}$ in this case, determines the interference structure. Returning to the OB example, the magnetic field is so weak that its effect is not  significant anymore. Instead, the electric field component of the laser can cause the electron displacement. As a result, the displacement must be in the same direction as the laser polarization. That is why the direction of the interference structure is different from that in XUV case. Let's look at the process of the ionization. Due to the strong laser intensity in the OB ionization, the electron can be ionized before the peak of the electric field. In a deep OB regime such as the case we are discussing, the peak intensity is so high that we can think the majority of electron wave packet escapes from the nucleus at the onset of the laser pulse approximatively. The electron's velocity is $\bm{A}_{\rm OB}(t)$ if the Coulomb force is ignored in this procedure. At the end of the laser pulse, the electron's velocity becomes zero and the electron displacement is $\int_{-\infty}^{\infty}\bm{A}_{\rm OB}(t)\mathop{}\!\mathrm{d} t$. For the particular pulse parameters used here, the displacement is identical in amplitude to that of XUV case but along different directions. According to our Fresnel diffraction picture, they will present similar patterns in the PMDs. As seen in Fig.\,1, this is indeed the case.

\vskip 4mm
\centerline{\includegraphics[width=\linewidth]{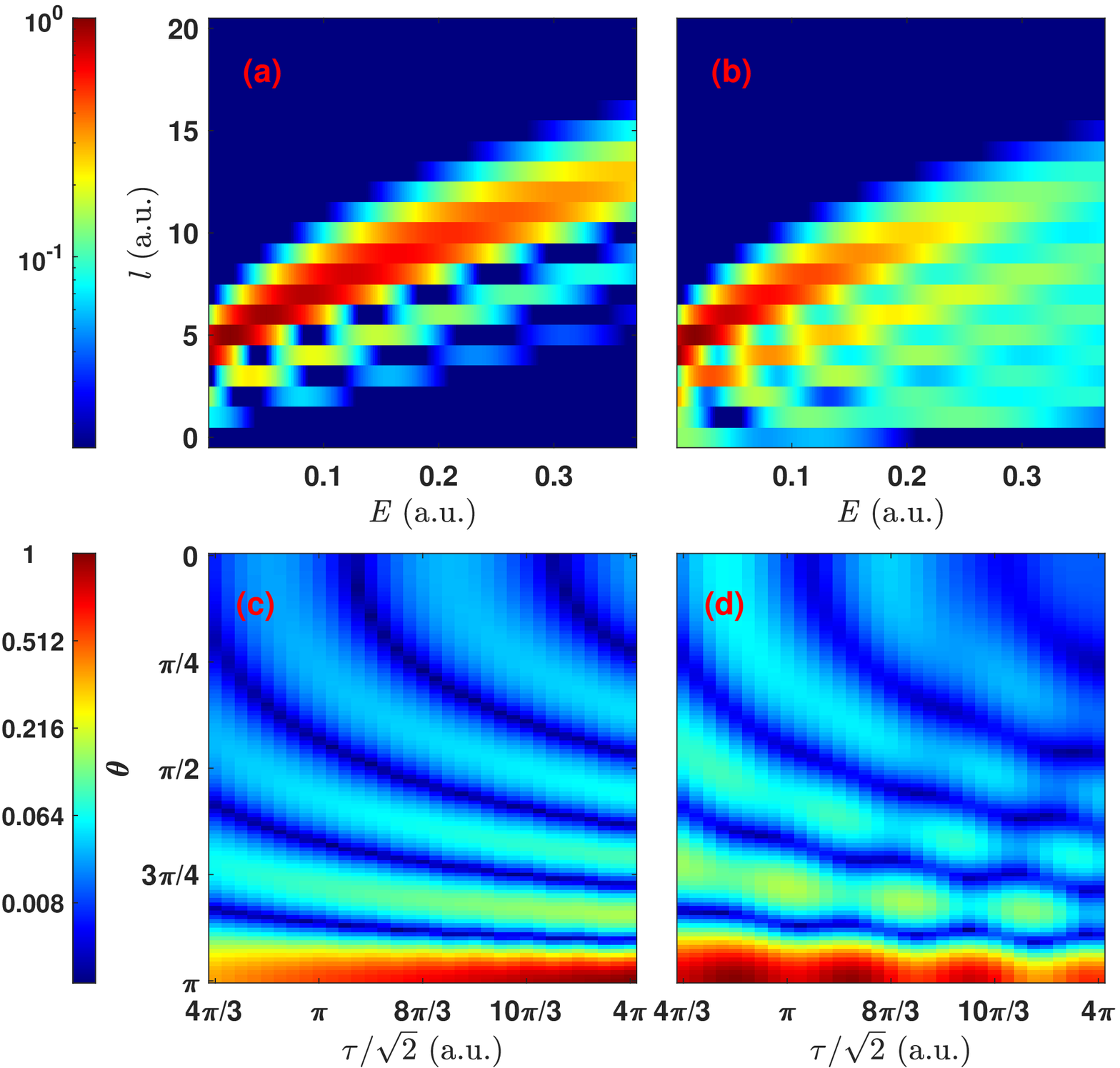}}
\vskip 2mm
\centerline{\footnotesize \begin{tabular}{p{7.5cm}}\bf Fig.\,2. \rm
    Color mappings of the energy-angular momentum distribution calculated from the TDSE for XUV case~(a) and OB case~(b). And the normalized angular distributions at $E_s = 10^{-4} {\rm a.u.}$ for different FWHMs $\tau$ are shown.
\end{tabular}}
\vskip 0.5\baselineskip

In Ref.~\cite{Geng2021}, we demonstrated the energy-angular momentum distributions and zero-energy angular distributions for different pulse FWHMs. To validate the correctness of the Fresnel diffraction picture in the OB condition, these two distributions for  OB cases are compared with those of XUV cases, which are presented in Fig.\,2. On the left is for the XUV case, while on the right is for the OB case. They still show comparable patterns. It should be noted that when changing FWHMs, the XUV laser's peak intensity remains constant, however, this is only true for the maximal vector potential in OB case.  The actual intensity and the effective frequency in the OB case will change according to its definition. For the OB case, in each arm of the  interference structure in Fig.\,2\,(d),  there exist additional structures caused by interferences  between  the ionization events at  different peaks of the electric field. Though the majority of electron wave packet is ionized at the start of the laser pulse, there is still minority remaining to be ionized at the second peak of the electric field. We can find that the contrast increases with the increasing of $\tau$. This is because that a larger FWHM will lead to a lower peak intensity of OB case, and a lower peak intensity will result in more bursts in the second peak, which indicates a larger interference contrast. The results of XUV cases have been predicted well by our quantum and semiclassical models, thus the same is true for results of OB cases.

Now let's concentrate on the tunneling scenario. In this section, a half-cycle pulse with a center wavelength 800~nm is used. Its vector potential can be written as
\begin{equation}
    {\bm A}_{\rm TU}(t)=\frac{\sqrt{I_{\rm TU}}}{\omega_{\rm TU}} \exp\left(-4\ln2\frac{t^2}{\tau_{\rm TU}^2}\right) \sin(\omega_{\rm TU} t)~{\hat{\bm e}_x},
    \label{laser}
\end{equation}
where $I_{\rm TU}$ is the peak intensity, $\omega_{\rm TU}$  is the carrier frequency and $\tau_{\rm TU}$ represents the FWHM. The laser pulse parameters are as follows: $\tau_{\rm TU}=\sqrt{2}\pi/2\omega_{\rm TU}\approx 39~{\rm a.u.}$ and $I_{\rm TU}=3\times10^{14}~{\rm W/{cm}^2}$, which means the Keldysh parameter is roughly 0.615. Fig.\,3\,(a) shows the vector potential and the electric field, while Fig.\,3\,(b) shows the corresponding PMD. Although there are some different structures in the high-energy region, the PMD also has a petal-like structure in the low-energy region, which implies that the underlying mechanism may be similar to previous cases.

\vskip 4mm
\centerline{\includegraphics[width=\linewidth]{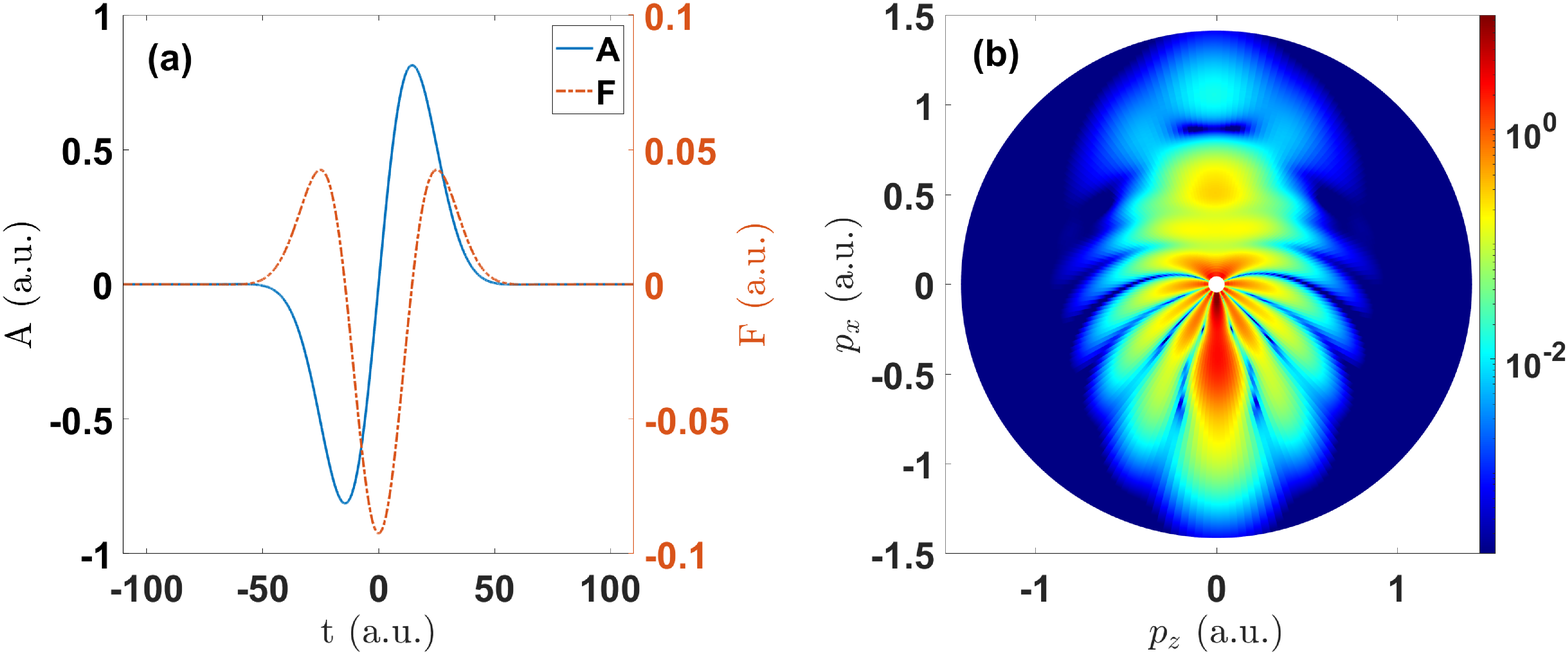}}
\vskip 2mm
\centerline{\footnotesize \begin{tabular}{p{7.5cm}}\bf Fig.\,3. \rm
    The vector potential, electric field and the PMD in the $xz$-plane for the tunneling case. See the main text for laser parameters.
\end{tabular}}
\vskip 0.5\baselineskip

\begin{figure*}[htb]
\vskip 4mm
\centerline{\includegraphics[width=\linewidth]{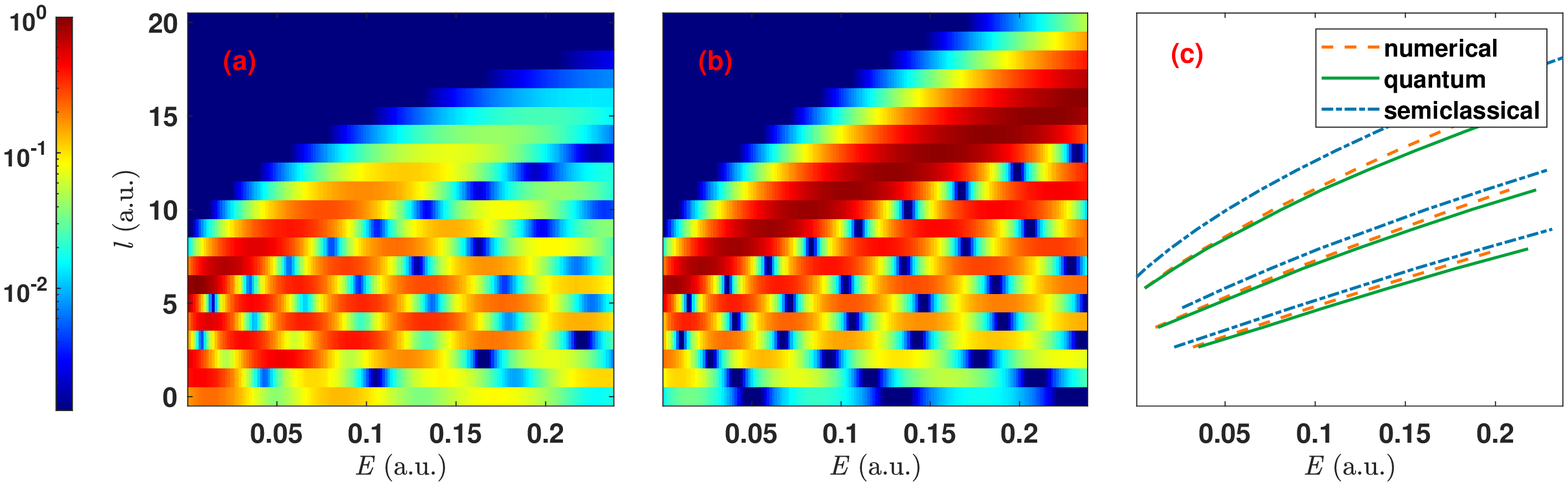}}
\vskip 2mm
\centerline{\footnotesize \begin{tabular}{p{15cm}}\bf Fig.\,4. \rm
    Color mappings of the energy-angular momentum distribution calculated from the TDSE (a) and quantum model (b). The comparison of the peak positions of these results is shown in (c), against the semiclassical predictions for n = 0, 1, 2 from the top to the bottom, respectively.
\end{tabular}}
\vskip 0.5\baselineskip
\end{figure*}

In this case, the peak intensity is even lower compared with the preceding examples. It is in the tunneling regime according to the Keldysh theory~\cite{keldysh1965} and OB criterion~\cite{Morishita2013}. After the laser pulse, the majority of the ground wave function remains unionized. Meanwhile, the majority of photoelectron is ionized when electric field reaches the maximum, which coincides with the time of $A=0$($t=0$) in Fig.\,3\,(a). It implies that the majority of photoelectrons have a velocity of zero at the end of the pulse. Then the following process will be dominated by electron diffusion. From the perspective of Fresnel diffraction, we should analyze the electron displacement in the ionization process first. After the electron tunnels from the atom, the canonical momentum of the electron remains conserved if the Coulomb force of the nucleus is ignored. The electron will have a velocity of $\bm{A}_{\rm TU}(t)$ at time $t$ after its tunneling around $t=0$. This leads to an accumulated electron displacement of $\int_{0}^{\infty}\bm{A}_{\rm TU}(t) \mathop{}\!\mathrm{d} t$ at the end of the laser pulse. In addition, the initial position~(the tunneling exit) of the electron is not at the nucleus, which can be estimated as $I_p/F$, where $I_p$ is the ionization energy and $F$ is the maximal electric field. It is critical to include this portion into the final displacement of the electron. The formula of the entire displacement can be written as
\begin{equation}
    \begin{aligned}
    z_0&=\frac{I_p}{F}+\int_0^{\infty} A_{\rm TU}(t) \mathop{}\!\mathrm{d} t\\
    &=\frac{I_p}{\sqrt{I_{\rm TU}}}+\frac{\sqrt{I_{\rm TU}}}{\omega_{\rm TU}^2}\times C\times {\rm {Dawson}}(\frac{C}{2}),
    \end{aligned}
\end{equation}
where $C$ is a constant $C=\pi/(2\sqrt{\ln4})$ and ${\rm {Dawson}}(x)$ is the Dawson integral.

After being calculated using the method above, the displacement in the tunneling case can be easily substituted into formulas of our quantum and semiclassical models to get ionization spectra. And then it can be compared with the results from TDSE. In Ref.~\cite{Geng2021}, the energy-angular momentum distribution of quantum model is represented as
\begin{equation}
    \left| a(E,l)\right|^2\propto\frac{2l+1}{\sqrt{E}}\left|R_{\sqrt{2E}l}(z_0)\right|^2,
    \label{a_l_e}
\end{equation}
where $R_{kl}(r)$ is the radial part of the scattering state. For the semiclassical model, the maximum position of the distribution can be given by~\cite{Geng2021}
\begin{equation}
    \int_{r_{\rm{min}}}^{z_0}\sqrt{2(E-\frac{l^2}{2r^2}+\frac{1}{r})}\mathop{}\!\mathrm{d} r=n\pi,
  \label{other_branches}
\end{equation}
where $r_{\rm{min}} =  {(\sqrt{1+2El^2}-1)}/{2E}$ and $n$ is a non-negative integer. All the results are shown in Fig.\,4. The interference structure of {\it ab initio} calculation can be reproduced well by our models with modified displacement $z_0$.

To confirm the general validity of our models in the tunneling regime, numerical computations under different laser intensities have been carried out. The peak intensity is increased from $1\times10^{14}{\rm W/{cm}^2}$ to $5.4\times10^{14}{\rm W/{cm}^2}$. The normalized photoelectron angular distributions at $E_s=10^{-4}\,{\rm a.u.}$ from the TDSE are shown in Fig.\,5\,(a). It's important to notice that the overall ionization rate increases when the peak intensity increases. According to Ref.~\cite{ammosov1987}, the tunneling rate of the atomic hydrogen's ground state is given by
\begin{equation}
    W(F)=\frac{4}{F}\exp\left(-\frac{2}{3F}\right),
    \label{ionization_rate}
\end{equation}
where $F$ is the electric field. When the peak intensity is changed, the factor need to be multiplied into the previous formulas of the quantum model to acquire the amended formula~\cite{Geng2021,Landau1981}
\begin{equation}
    \left| a(\theta,z_0,F) \right|^2\propto W(F)\left| \sum_{l=0}^{l_{\max}} \frac{2l+1}{\sqrt{z_0}}J_{2l+1}(\sqrt{8z_0})P_l(\cos\theta)\right|^2,
    \label{a_quantum}
\end{equation}
where $P_l$ is the Legendre polynomial, $J$ is the Bessel function and $l_{\max}$ is set to 25 in the present calculations. The results of the quantum model are shown in Fig.\,5\,(b). It is clear that the TDSE result in Fig.\,5\,(a) exhibits additional slowly varying interference structures compared with that of the quantum model. They are caused by the interference between the ionization at the main peak and those at the other two sub-peaks, which is not taken into account in our simplified models. Otherwise, the position of the interference peak is provided by $4\sqrt{2z_0}\sin({\theta}/{2})=2n\pi$ if using the semiclassical model~\cite{Geng2021}. For a better comparison, the peak positions of distributions from both TDSE and the quantum model are plotted against those of the semiclassical model, as shown in Fig.\,5\,(c). Despite the presence of ionization at two sub-peaks, the numerical results are in good agreement with the two models' results.

\begin{figure*}[htb]
\vskip 4mm
\centerline{\includegraphics[width=\linewidth]{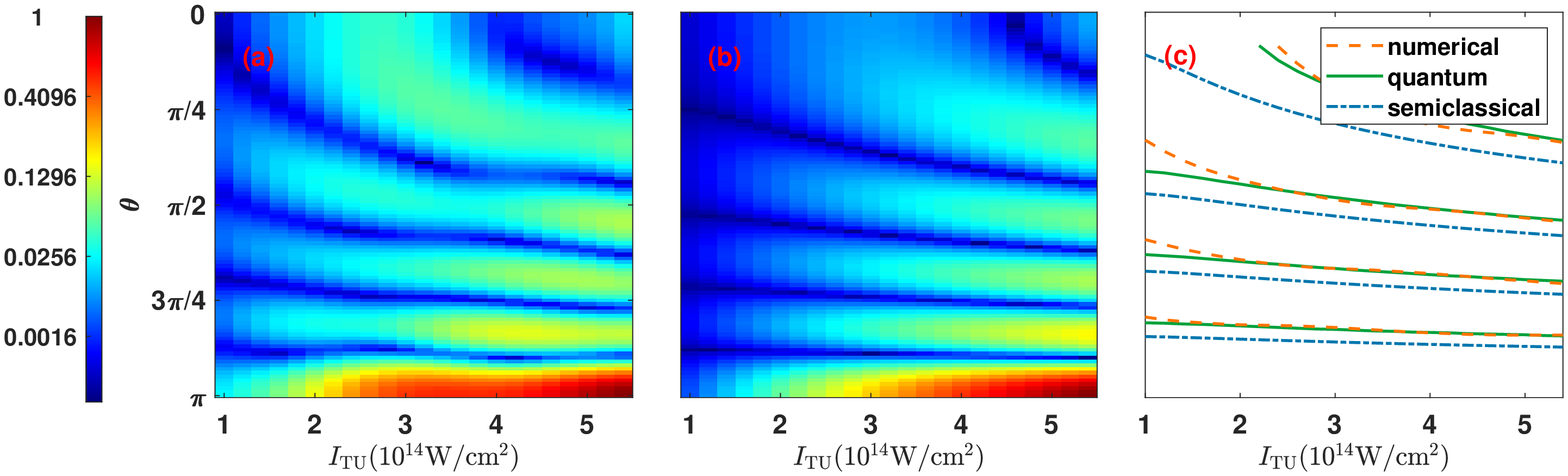}}
\vskip 2mm
\centerline{\footnotesize \begin{tabular}{p{15 cm}}\bf Fig.\,5. \rm
    The normalized angular distributions at $E_s=10^{-4}\,{\rm a.u.}$ calculated from the TDSE (a) and quantum model (b) for different laser intensities. The comparison of the peak positions of these results is shown in (c), against the semiclassical predictions.
\end{tabular}}
\vskip 0.5\baselineskip
\end{figure*}

The ionization in half-cycle or single-cycle pulses attracts theorists' interest all the time~\cite{yuan2021,manescu2003}. It helps us understand the sub-cycle ionization dynamics. In this Letter, we investigate this topic from an unusual perspective, which is the Fresnel diffraction picture developed in our prior work on super-intense XUV. The picture illustrates similarities of low-energy photoelectron behaviors in the strong field. The entire ionization process can be divided into three stages. At the beginning, the electron is bounded by the Coulomb field. After being ionized and during the laser pulse, the electron is driven by the external field. When the laser pulse ends, the electron has a velocity close to zero and is controlled by the Coulomb field again. A laser-induced displacement in the second stage can be calculated and then be inserted into the analytical formulas that represents the third stage to achieve the final results, provided that the displacement is remarkably larger than the atom radius. Our picture is ideal for the condition that the velocity of the main photoelectron wavepacket is roughly equal to zero at the end of the pulse. By contrast, the common direct ionization and recollision picture will have difficulty in dealing with such instances. We just show two typical half-cycle cases that can be solved concisely and elegantly in this Letter. It should be mentioned that the recollision process is not taken into account in our model. The low-energy structures will be smeared out if we use the long-duration laser or half-cycle pulses with other carrier envelope phases due to the interference caused by recollisions. And the multi-cycle effects also disrupt the structure for long-duration cases. In the experiments, mid-infrared laser technology has advanced dramatically over recent years. A method for producing powerful optical half-cycle attosecond pulses was devised and tested~\cite{Wu2012,Xu2018}. The laser parameters used in this Letter are within the equipment's capability, which  provides the possibility  to test the present finding. Furthermore, when dealing with more complex systems such as molecules, the structure of the system also has an impact on the result of the diffraction.  While beyond the scope of this paper, it would be an interesting topic to investigate the structure information reflected in the low-energy PMD.

In conclusion, we have studied the dynamics of low-energy photoelectrons interacting with the half-cycle pulse in the tunneling and OB regime. It demonstrates that the low-energy structure is similar to that in our previous work on super-intense XUV. This can be attributed to the fact that the  underlying physical mechanisms  are  quite similar. The quantum and semiclassical models  after appropriate modifications can reproduce the  results of the TDSE.  It highlights the generality of our laser-induced Fresnel diffraction picture and provides an intuitive picture for the ionization dynamics of low-energy electrons without recollisions.

This work is supported by the National Natural Science Foundation of China~(NSFC) under Grant Nos.~11961131008 and 11725416, by the National Key R\&D Program of China under Grant No. 2018YFA0306302, and by the National Science Centre (Poland) under Grant No. 2018/30/Q/ST2/00236.

\bibliography{mybibtex}
\end{document}